\documentclass[a4paper,fleqn,usenatbib]{mn2e}
\usepackage[T1]{fontenc}
\usepackage{ae,aecompl}

\usepackage[latin1]{inputenc}
\usepackage{graphicx}   
\usepackage{amsmath}
\usepackage{amsfonts}
\usepackage{amssymb}
\usepackage{color}

\usepackage{times}

\newcommand{\Msun}{\ensuremath{\textrm{M}_{\odot}}}

\newcommand{\Lsun}{\ensuremath{\textrm{L}_{\odot}}}

\newcommand{\kms}{km\hspace{0.25em}s$^{-1}$}

\newcommand{\HI}{\mbox{H\hspace{0.25em}{\sc i}}}

\newcommand{\HeI}{\mbox{He\hspace{0.25em}{\sc i}}}
\newcommand{\HeII}{\mbox{He\hspace{0.25em}{\sc ii}}}

\newcommand{\OI}{\mbox{O\hspace{0.25em}{\sc i}}}
\newcommand{\OII}{\mbox{O\hspace{0.25em}{\sc ii}}}
\newcommand{\CI}{\mbox{C\hspace{0.25em}{\sc i}}}
\newcommand{\CII}{\mbox{C\hspace{0.25em}{\sc ii}}}
\newcommand{\CIII}{\mbox{C\hspace{0.25em}{\sc iii}}}

\newcommand{\MgII}{\mbox{Mg\hspace{0.25em}{\sc ii}}}

\newcommand{\AlIII}{\mbox{Al\hspace{0.25em}{\sc iii}}}

\newcommand{\SiII}{\mbox{Si\hspace{0.25em}{\sc ii}}}
\newcommand{\SiIII}{\mbox{Si\hspace{0.25em}{\sc iii}}}
\newcommand{\SiIV}{\mbox{Si\hspace{0.25em}{\sc iv}}}
\newcommand{\CaII}{\mbox{Ca\hspace{0.25em}{\sc ii}}}

\newcommand{\TiIII}{\mbox{Ti\hspace{0.25em}{\sc iii}}}

\newcommand{\FeII}{\mbox{Fe\hspace{0.25em}{\sc ii}}}
\newcommand{\FeIII}{\mbox{Fe\hspace{0.25em}{\sc iii}}}

\newcommand{\CoIII}{\mbox{Co\hspace{0.25em}{\sc iii}}}

\newcommand{\Cofs}{$^{56}$Co}
\newcommand{\Nifs}{$^{56}$Ni}

\newcommand{\Mej}{M$_{\textrm{ej}}$}

\newcommand{\KE}{\ensuremath{E_{\rm K}}}

\newcommand{\aap}{A\&A}

\newcommand{\apj}{ApJ}
\newcommand{\apjs}{ApJS}
\newcommand{\apjl}{ApJ}
\newcommand{\aj}{AJ}

\newcommand{\mnras}{MNRAS}

\newcommand{\nat}{Nature}

\newcommand{\araa}{ARA\&A}

\newcommand{\eg}{e.g.,\ }

\def\gsim{\mathrel{\rlap{\lower 4pt \hbox{\hskip 1pt $\sim$}}\raise 1pt \hbox {$>$}}}
\def\lsim{\mathrel{\rlap{\lower 4pt \hbox{\hskip 1pt $\sim$}}\raise 1pt \hbox {$<$}}}

\definecolor{myorange}{rgb}{0.9,0.6,0.0}

\definecolor{myredder}{rgb}{1.0,0.0,0.0}

\definecolor{mygreen}{rgb}{0.0,0.7,0.0}

\definecolor{myblue}{rgb}{0.0,0.0,1.0}

\definecolor{myyellow}{rgb}{0.92,0.89,0.00}

\begin{document}

\title[Spectrum formation in SLSNe]
  {Spectrum formation in Superluminous Supernovae (Type I)}
\author[P. A.~Mazzali et al.]{P. A. Mazzali$^{1,2}$
\thanks{E-mail: P.Mazzali@ljmu.ac.uk}, 
M. Sullivan$^{3}$, 
E. Pian$^{4,5}$,
J. Greiner$^{6}$,
D. A. Kann$^{7}$
\\
\\
  $^{1}$Astrophysics Research Institute, Liverpool John Moores University, IC2, Liverpool Science Park, 146 Brownlow Hill, Liverpool L3 5RF, UK\\
  $^{2}$Max-Planck-Institut für Astrophysik, Karl-Schwarzschild-Str. 1, D-85748 Garching, Germany\\
  $^{3}$School of Physics and Astronomy, University of Southampton, Southampton, SO17 1BJ, UK\\
  $^{4}$Institute of Space Astrophysics and Cosmic Physics, via P. Gobetti 101, 40129 Bologna, Italy \\
  $^{5}$Scuola Normale Superiore, Piazza dei Cavalieri 7, I-56126 Pisa, Italy \\
  $^{6}$Max-Planck-Institut für Extraterrestrische Physik, Giessenbach-Str. 1, D-85748 Garching, Germany\\
  $^{7}$Th\"uringer Landessternwarte Tautenburg, Sternwarte 5, 07778 Tautenburg, Germany 
}

\date{Accepted ... Received ...; in original form ...}
\pubyear{2015}
\volume{}
\pagerange{}

\maketitle

\begin{abstract}  
The near-maximum spectra of most superluminous supernovae that are not dominated
by interaction with a H-rich CSM (SLSN-I) are characterised by a blue spectral
peak and a series of absorption lines which have been identified as \OII.
SN\,2011kl, associated with the ultra-long gamma-ray burst GRB111209A,  also had
a blue peak but a featureless optical/UV spectrum. 
Radiation transport methods are used to show that the spectra (not including 
SN\,2007bi, which has a redder spectrum at peak, like ordinary SNe\,Ic) can be 
explained by a rather steep density distribution of the ejecta, whose 
composition appears to be typical of carbon-oxygen cores of massive stars which
can have low metal content. 
If the photospheric velocity is $\sim 10000-15000$\,\kms\, several lines form
in the UV. \OII\ lines, however, arise from very highly excited lower levels, 
which require significant departures from Local Thermodynamic Equilibrium to be 
populated. These SLSNe are not thought to be powered primarily by \Nifs\ decay. 
An appealing scenario is that they are energised by X-rays from the shock 
driven by a magnetar wind into the SN ejecta.
The apparent lack of evolution of line velocity with time that characterises
SLSNe up to about maximum is another argument in favour of the magnetar 
scenario. The smooth UV continuum of SN\,2011kl requires higher ejecta 
velocities ($\sim 20000$\,\kms): line blanketing leads to an almost featureless 
spectrum. 
Helium is observed in some SLSNe after maximum. The high ionization near maximum
implies that both He and H may be present but not observed at early times.  
The spectroscopic classification of SLSNe should probably reflect that of 
SNe\,Ib/c. Extensive time coverage is required for an accurate classification.   
\end{abstract}


\section{Introduction}
\label{sec:introduction}

Supernovae (SNe) are physically divided into two groups, core-collapse and
thermonuclear. Classically, they are typed based on their light curves and
spectra \citep[see][for a review]{filippenko97}. Recently, however, new classes
of SNe have emerged that only partially fit into this classical scheme. In
particular, a number of very luminous events have been detected by different
surveys, many at redshifts exceeding 0.5 \citep[e.g.,][see \citet{GalYam12} for
a
review]{quimby07,barbary09,quimby11,chomiuk11,leloudas12,lunnan13,howell13,papadopoulos15}.
Although they are collectively classified as superluminous supernovae (SLSNe),
their observational properties are quite diverse.

Some SLSNe clearly bear the sign of interaction with a hydrogen-rich
circum-stellar medium \citep[CSM;][]{smith07,ofek07,drake11,benetti14}. This
masks the actual SN ejecta, making their nature unclear, although a link with
massive stars is likely \citep[\eg][]{smith07,agnoletto09}.  SLSNe without
the signatures of interaction with a H-rich CSM, (``H-poor SLSNe'') are in turn 
divided into two rather vaguely defined groups.
Most of them have very blue spectra at maximum which show strong \OII\ lines
\citep[SLSN-I,][]{GalYam09}.
Some, such as SN2007bi \citep{GalYam09}, have redder maxium-light spectra,
closer to those of type Ic SNe (SNe\,Ic), and are pair-instability SN (PISN) 
candidates.  Both of these latter groups would technically be classified as 
SNe\,Ic, or at least as stripped-envelope events, since hydrogen lines are 
never strong. Finally, the ULGRB\,111209A/SN\,2011kl is intermediate in 
luminosity between GRB/SNe and SLSNe, but its spectral energy distribution 
resembles that of SLSNe.

Various interpretations have been proposed for these SNe. 
The PISN candidate SN\,2007bi \citep{GalYam09} has very extended optical light 
curves, and emits strong forbidden lines of [\FeII] at late times. This is 
consistent with a \Nifs-driven light curve, and suggestive of the explosion of 
a very massive star with a zero-age main sequence mass close to, or in excess 
of 100\,\Msun. This scenario has been criticized \citep[\eg][]{nicholl13}, but 
in our opinion the presence of forbidden emission lines of Fe in an amount
compatible with the \Nifs\ that is necessary to drive the light curve
\citep{GalYam09} makes the Ni-driven scenario uncontroversial. The massive
collapse scenario \citep{moriya10} agrees on this despite uncertainties. 
For none of the other H-poor SLSNe has the presence or amount of \Nifs\ 
been confirmed. These SNe typically show strong absorption lines of lighter 
elements, including oxygen, and have never been observed in the nebular phase 
as they are typically too distant to be observed at late times.  
Their light curves do not appear to be consistent with being powered by \Nifs\ 
decay \citep{inserra13,howell13,papadopoulos15}.  
One alternative to \Nifs\ is that the SN is powered by energy emitted by a
magnetized, rapidly spinning neutron star (a Magnetar), 
\citep[\eg][]{Bucciantini09}. In a magnetar-powered scenario interaction of a 
magnetar-driven shock with the SN ejecta could transform kinetic energy into 
radiative energy and power the SN light curve, leading to SLSNe 
\citep{kasen10,woosley10}. 
Another possibility is interaction with H- and He-free CSM 
\citep{sorokina15,moriya12,chatzwhee12,baklanov15}. These studies have focussed
on reproducing the light curves of SLSNe. However, spectra should also be
explained. One signature of interaction with an H- and He-free CSM would be 
shallow absorption lines, and a spectrum that looks too cool for its luminosity,
which is the sum of two components, the SN and the interaction. An example is 
the "Super-Chandrasekhar" SN\,Ia 2009dc \citep{hachinger12}. This is not what 
we see in SLSNe. We see the opposite: a spectrum which is too hot for its
luminosity. The presence of an outer CSM does not seem to be compatible with
the  spectra of this class of SLSNe, which we therefore call
``non-interacting''.

Useful information about SN ejecta can be obtained from modelling their spectra.
As the SN ejecta expand, deeper and deeper layers of the exploded star are
exposed, and spectra reveal their properties. A consistent calculation can be
used to determine the abundances in the ejecta, and these in turn may indicate
the properties of the progenitor and the effect of the explosion. This has been
demonstrated for type Ia SNe \citep[\eg ][]{mazzali08}, as well as for SNe Ib/c 
\citep[\eg ][]{sauer06}. However, for SLSNe the only attempts to model the
spectra have relied on parametrised codes, which are effective in suggesting
line identifications but cannot provide quantitative information with respect to
masses, energies and abundances \citep[\eg][]{vreeswijk14}. Here we apply our
models to SLSNe. Our methodology is outlined in Section \ref{sec:method}.

A variety of SLSN spectra have been presented in various papers
\citep[e.g.,][]{quimby11,chomiuk11,inserra13,vreeswijk14,nicholl14}. Only in a
few cases are time-series of spectra available. We have selected iPTF13ajg
(Section \ref{sec:13ajg}) and PTF09cnd (Section \ref{sec:09cnd}) as typical
examples of line-rich SLSNe. We also examine SNLS-06D4eu  \citep{howell13}, the
most distant spectroscopically confirmed published  SLSN to date (Section
\ref{sec:06D4eu}).  We then compare the results with a line-poor, luminous 
SN\,2011kl/GRB111209A \citep{greiner15} in Section~\ref{sec:SN2011kl}.  We
conclude with a discussion of the general properties of SLSN spectra, and the
inferences that can be made about these events from their spectral properties
(Section \ref{sec:disc}).

\section{Method}
\label{sec:method}

The spectra of SNe (and SNe Ic in particular) can have lines of different
widths. SNe Ic are a particularly interesting case, as the range goes from
narrow \citep[a few 1000\,\kms, as for SN 1994I, ][]{sauer06} to very broad
lines \citep[$\sim 60000$\,\kms, as for SN 2002ap, ][]{mazzali02}. This is
roughly reflected in the classification of SNe Ic versus broad-lined (BL) SNe
Ic. Modelling is necessary to turn the intrinsic properties of the gas in the SN
ejecta (density and abundance distributions, excitation and ionization degree
and their radial dependence) into observed line widths, taking into account the
effect of line blending, which is particularly important in the rapidly
expanding SN ejecta where the material follows a Hubble-type law.  

The spectra of SLSNe are characterised by lines of mostly singly ionized
elements, but are unique among SNe in showing \OII\ lines.  Line identification,
while probably correct especially when multiple features are identified, is
often performed using codes that do not treat the level occupation numbers
and the radiation field consistently. We therefore need to verify these results.
Interestingly, line velocity in SLSNe does not seem to evolve much with time
\citep[see, \eg][]{chomiuk11,nicholl14,vreeswijk14}. This is different from the
typical behaviour of SN spectra. The question is what model can reproduce this. 

We modelled the SLSN spectra using our Monte Carlo supernova spectrum synthesis
code. This has been described in several papers \citep{ml93,m00} and has been
used to model SNe of different spectral types \citep{m13,mazzali14b}. The code
assumes that the SN luminosity is emitted at a lower boundary as a black body
and transported through an expanding SN `atmosphere' of variable density and
composition. The atmosphere is assumed to be in radiative equilibrium, and
radiation can be absorbed in lines or scattered off electrons. Level populations
are computed using a modified nebular approximation, which is appropriate for
the conditions prevailing in SN ejecta \citep{pauldrach96}. The emergent
spectrum is computed both by counting emerging energy packets (which are used as
proxies for photons), and by solving the formal integral of the transfer
equation using as source functions and opacities obtained in the Monte Carlo
experiment by iterating with the radiation field \citep{lucy99}. Because our
code solves for the level populations and the radiation field consistently, it
can be used to estimate masses and abundances in the outer part of the SN ejecta
\citep[see, \eg][]{mazzali14a}.

While the code is not fully self-consistent, in that it does not depend on a
specific explosion model but rather uses one to describe the layers above a
sharp photosphere, it is also not parametrized, because occupation numbers, and
hence line opacities, are computed consistently with the properties of the
radiation field. It therefore goes beyond simple line identification, making it
possible to quantify relative and absolute elemental abundances simultaneously
with the density and mass in the expanding SN ejecta. With the method of
abundance tomography \citep[\eg][]{stehle05} it is possible to take advantage of
the progressive thinning out of the SN ejecta and explore the properties of
deeper and deeper layers.  

The code can treat non-thermal excitation of He consistently, computing the
effect on the level population of non-thermal particles as \eg\ produced by
radioactive decays \citep{lucy91,ml98,hachinger12}, if the radiation field is
known. This is possible for SNe\,Ib because we know that the non-thermal
radiation field comes from the incomplete thermalization of the gamma-rays and
positrons produced by the decay of \Nifs\ and \Cofs. As an alternative,
non-thermal effects can be simulated in a parametrised way, modifying the
population of the excited levels of relevant ions by introducing departure
coefficients that should imitate a consistent calculation
\citep[\eg][]{mazzali09}. We use the latter approach here because we do not know
the details of the non-thermal radiation field and do not treat non-thermal
effects in \OII\ explicitly. However, since the \OII\ ion is characterised by
excited levels with a very high excitation energy, it can be expected to be
sensitive to non-thermal radiation in a way similar to \HeI. We therefore apply
departure coefficients similar to those obtained for \HeI\ 
\citep{lucy91,ml98,hachinger12}, mimicking non-thermal effects on \OII.

\section{Line-rich SLSNe: iPTF13ajg }
\label{sec:13ajg}

We begin with the spectra of one of the best observed SLSNe, iPTF13ajg, at
redshift $z=0.74$. This supernova was discovered by the intermediate Palomar
Transient Factory (iPTF). A large dataset for this event was presented in
\citet{vreeswijk14}, making this an excellent `template' SLSN, since its
behaviour is similar to that of many other less well observed cases. The spectra
are characterised by a slow evolution.  The strongest lines have been identified
as \FeIII, \CII, \CIII, \SiIII, \OII\ \citep{vreeswijk14}. The lines are broad,
but not extremely broad. A characteristic line width seems to be $\sim
15000$\,\kms. Line blending is not as dominant as in GRB/SNe, for example
\citep[see, \eg][]{iwamoto98}, or even in non-GRB BL-Ic SNe \citep[\eg][]{m13}.
However, line width is not easy to relate to a velocity spread, as most lines
are actually blends. Modelling is necessary to address this. 

The lines that have been identified are not indicative of a particularly high
temperature, despite the high luminosity. This is understandable, because SLSNe
evolve slowly in luminosity, and therefore they are observed with much larger
radii than SNe\,Ib/c, which reduces the temperature. The only indication of a
very high temperature is the presence of \OII\ lines. It has been noted that the
ionization potential of \OI\ is high. While this is true, it is similar to \HI\
(13.6 eV), and much lower than that of \HeI\ (24.6 eV).  However, this is not
the whole story. The observed lines, if they are correctly identified as \OII,
arise from very highly-excited lower levels, which have excitation potentials of
$\sim 25 $\,eV. This is higher than the energy required to excite \HeI\ lines in
SNe\,Ib ($\sim 20 $\,eV), and indeed comparable to the ionization potential of
\HeI. It is therefore to be expected that the excitation of \OII\ levels is not
in thermal equilibrium with the local radiation field. We can test this with our
models, and attempt to derive relative abundances in the SN ejecta. This should
in turn shed light on the nature of the progenitor and on the mechanism that
powers SLSNe.

Line velocity evolves very slowly with time in iPTF13ajg, as in other SLSNe.  In
order for the position of the lines not to move significantly over several weeks
or months despite quite a high expansion velocity, either a steady-state
situation holds (as in stellar winds), or the density profile must be very
steep.  In the former case, we would expect very broad lines, since $\rho
\propto r^{-2}$, which leads to significant mass being able to absorb radiation
over a large range of velocities.  Furthermore, the spectra of SLSNe (iPTF13ajg
is very well-observed but others show a similar behaviour) are seen to become
redder with time as the luminosity evolves. This is typical of SNe and would not
be expected in a steady-state regime. We therefore explore the latter option.

In order to reproduce the spectra of iPTF13ajg we used a steep density profile,
$\rho \propto r^{-7}$, which is typical of low-energy explosions of compact
stars \citep[\eg][]{sauer06}. Line width depends on the gradient of the density
\citep[see, \eg][]{mazzali00}. Other characteristic features of SLSNe are that
they are extremely luminous, but also rather long-lived. In the case of
iPTF13ajg the models we tested have ejecta mass ranging from $\sim 25$ to $\sim
50$\,\Msun. The mass is always quite large mass, which is required by the long
risetime, the high luminosity and the relatively large velocity
\citep{vreeswijk14}.

\subsection{Pre-Maximum epochs}
\label{sec:pre-max}

The spectra of PTF13ajg and similar SLSNe before and around maximum light are
characterized by strong absorption lines, both in the rest-frame ultraviolet
(UV) and in the optical region. The spectra typically turn over at $\sim
2700$\,\AA. This is extremely blue for any SN: normally SN spectra are subject
to strong metal absorption (Fe, Co, Ni, Ti, Cr) shortwards of $\sim 4000$\,\AA.
The behaviour of SLSNe may indicate a higher temperature than in other
SNe\,Ib/c, but this is in conflict with the ionization degree, which is not
extremely high, and with the fact that more highly ionized metal ions are even
more likely to absorb in the UV. It seems more likely that it is the result of a
low metal abundance.

\begin{figure}
 \includegraphics[width=89mm]{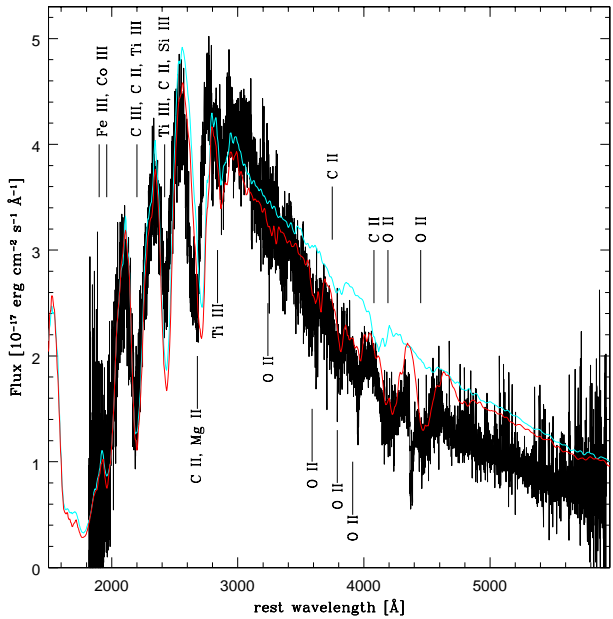}
\caption{The spectrum of iPTF13ajg on 9 April 2013, compared to two synthetic
spectra: one does not include an increase in the occupation numbers of \OII\ and
\HeI\ to account for non-thermal effects (cyan line), the other does (red
line).}
\label{fig:13ajgUV}
\end{figure}

We focus first on one of the best observed spectra of iPTF13ajg, obtained on
2013 April 9, about 8 days before maximum \citep{vreeswijk14}. We adopted a
distance modulus $\mu = 43.25$\,mag and no reddening. We used a rest-frame epoch
of 50 days for this spectrum, starting from the bolometric light curve shown in 
\citet[][figure 5]{vreeswijk14}, but we also tested a shorter risetime (40
days). This range of risetimes does not change the nature of our results but
does affect the estimate of the mass, as the density profile can be rescaled
down in mass, without altering the ratio between kinetic energy and ejected
mass. Fig.\,\ref{fig:13ajgUV} shows the spectrum along with two models. One
model was computed without considering non-thermal ionization/excitation, the
other includes it in a parametrized way. The models are characterized by the
same luminosity ($7 \times 10^{10}$\,\Lsun), photospheric velocity
(12250\,\kms), and abundances. The photospheric velocity does not evolve much
during this phase, and the mass that is responsible for line formation is only
$\sim 3$\,\Msun.  The composition is dominated by carbon (40 per cent by mass),
oxygen (48 per cent), helium (10 per cent), neon (2 per cent). Heavier elements
have low abundances (see Table \ref{tab:13ajg}). The  abundances were adjusted
to optimise the match to the observed spectrum, and they indicate substantially
sub-solar values ($\sim 1/4$ solar), based on the iron content.

\begin{table*}
\caption{Parameters of the models for iPTF13ajg}
\label{tab:13ajg}
\centering
\begin{tabular}{ccccccccccccccccc}
Date & Epoch & Lum     & v$_{ph}$ & T$_B$ & He  & C    & O     & Ne   & Mg   & Si   & S    & Ca   & Ti   & Fe	& Co   & Ni  \\
 & [rf days] & erg s$^{-1}$ & km s$^{-1}$ & K & \multicolumn{11}{c}{relative abundances}       \\
\hline
2013 Apr 8 &  49 & 44.48 & 12250 & 12600 & 0.10 & 0.40 & 0.475 & 0.02 & 5e-4 & 2e-3 & 5e-4 & 5e-5 & 2e-5 & 5e-4 & 3e-4 & 1e-5 \\
2013 Apr 9 &  50 & 44.44 & 12250 & 12100 & 0.10 & 0.40 & 0.475 & 0.02 & 5e-4 & 2e-3 & 5e-4 & 5e-5 & 2e-5 & 5e-4 & 3e-4 & 1e-5 \\
2013 Apr 16&  54 & 44.48 & 12250 & 11700 & 0.10 & 0.40 & 0.475 & 0.02 & 7e-4 & 2e-3 & 5e-4 & 5e-5 & 2e-5 & 5e-4 & 3e-4 & 1e-5 \\
2013 May 7 &  67 & 44.44 & 12250 &  9800 & 0.10 & 0.40 & 0.475 & 0.02 & 5e-4 & 2e-3 & 5e-4 & 5e-5 & 2e-5 & 5e-4 & 3e-4 & 1e-5 \\
2013 May 10&  69 & 44.44 & 11750 &  9900 & 0.10 & 0.40 & 0.475 & 0.02 & 5e-4 & 2e-3 & 5e-4 & 5e-5 & 1e-5 & 5e-4 & 3e-4 & 1e-5 \\
2013 Jun 6 &  84 & 44.28 & 11000 &  8500 & 0.10 & 0.40 & 0.475 & 0.02 & 3e-6 & 2e-3 & 5e-4 & 4e-5 & 2e-5 & 5e-4 & 2e-4 & 1e-5 \\
2013 Jun 10&  87 & 44.23 & 10750 &  8200 & 0.10 & 0.40 & 0.475 & 0.02 & 1e-5 & 2e-3 & 5e-4 & 4e-5 & 2e-5 & 5e-4 & 2e-4 & 1e-5 \\
2013 Jul 12& 106 & 43.81 &  7500 &  7900 & 0.10 & 0.40 & 0.475 & 0.02 & 3e-6 & 2e-3 & 5e-4 & 4e-5 & 7e-5 & 2e-4 & 8e-5 & 1e-5 \\
2013 Sep 9 & 141 & 43.58 &  7000 &  5400 & 0.10 & 0.40 & 0.475 & 0.02 & 1e-5 & 2e-3 & 5e-4 & 4e-5 & 1e-6 & 2e-4 & 1e-4 & 1e-5 \\
\hline
\end{tabular}
\end{table*}

\begin{figure}
 \includegraphics[width=89mm]{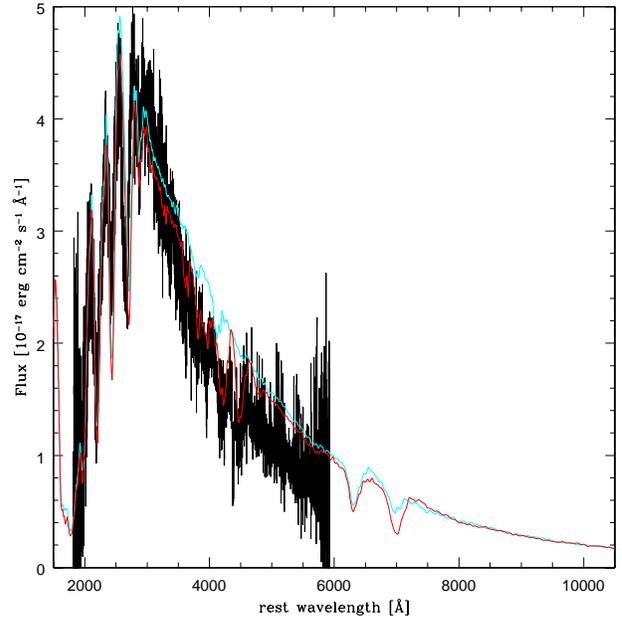}
\caption{As Fig.~\ref{fig:13ajgUV}, but showing the model prediction in the 
entire rest-frame optical domain.}
\label{fig:13ajgUVopt}
\end{figure}

The thermal model matches the observations reasonably well, but performs best
bluewards of 3000\,\AA. The main contributions to the various lines are marked
in Fig.~\ref{fig:13ajgUV}, and a list of the strongest lines is presented in
Table~\ref{tab:lines}. They are generally in agreement with previous results
\citep[\eg][]{vreeswijk14}, but with some differences. In particular, \TiIII\ 
leaves an important imprint on the spectrum, and it is solely responsible for
the line near 2900\,\AA. In the case of SN\,2011kl this is the only isolated
line that is seen \citep[][see Section~\ref{sec:SN2011kl}]{greiner15}. 
In the red part of the spectrum only weak \CII\ lines are produced by this
model, which may match some possible observed features but not the strongest
lines. These have typically been identified as \OII\ lines.

As mentioned above, all \OII\ lines come from highly excited lower levels. \OI\
itself has an ionization potential of 13.6\,eV, like \HI\ and similar to \FeII\
and \SiII. However, all of its NUV-optical lines arise from levels with
excitation potential ranging from 23 to 32\,eV. Thus, while it is reasonable to
expect that \OII\ and \FeIII\ are present, \OII\ lines should not be very strong
unless the temperature, and hence the ionization, is even higher. The strength
of the \OII\ lines cannot be increased by just increasing the oxygen abundance,
as this is already very high.

\begin{table}
\caption{The strongest lines in the spectra of iPTF13ajg and other similar SLSNe 
at pre-maximum epochs (the observed $\lambda$ refers to iPTF13ajg). Lines beyond
6000\,\AA\ are predicted by the model.}
\label{tab:lines}
\centering
\begin{tabular}{cccc}
feature $\lambda$ & ion & lab $\lambda$ & E$_l$    \\
  \AA    &        &   \AA      &   eV     \\
\hline
  1950   & \FeIII &  2061.56   &   5.08   \\
  	 & \FeIII &  2068.25   &   5.08   \\
	 & \FeIII &  2078.99   &   5.08   \\
	 & \CoIII &  2097.64   &   6.91   \\
  2200   & \CIII  &  2296.87   &  12.69   \\
  	 & \CII   &  2325.40   &   0.01   \\
	 & \TiIII &  2339.00   &   4.74   \\
  2400   & \CII   &  2512.06   &  13.72   \\ 
         & \TiIII &  2516.07   &   4.76   \\
         & \TiIII &  2527.85   &   4.74   \\
	 & \SiII  &  2541.82   &  10.28   \\
  2670   & \MgII  &  2795.53   &   0.00   \\
         & \CII   &  2836.71   &  11.96   \\
  2830   & \TiIII &  2984.74   &   5.17   \\	 
  3230   & \OII   &  3377.19   &  25.29   \\ 
  3600	 & \OII   &  3749.49   &  23.00   \\ 
  3800	 & \CII   &  3920.68   &  16.33   \\
         & \OII   &  3954.36   &  23.42   \\ 
  	 & \OII   &  3973.26   &  23.44   \\ 
  3900	 & \OII   &  4075.86   &  25.67   \\ 
  4200	 & \CII   &  4267.26   &  18.05   \\
         & \OII   &  4345.57   &  22.98   \\ 
  	 & \OII   &  4349.43   &  23.00   \\ 
  	 & \OII   &  4366.91   &  23.00   \\ 
  	 & \OII   &  4414.89   &  23.44   \\ 
  	 & \OII   &  4416.97   &  23.42   \\ 
  4450	 & \OII   &  4638.86   &  22.97   \\ 
         & \OII   &  4641.83   &  22.98   \\ 
         & \OII   &  4650.85   &  22.97   \\ 
         & \OII   &  4661.64   &  22.98   \\ 
  6250	 & \CII   &  6578.05   &  14.45   \\
  	 & \CII   &  6582.88   &  14.45   \\
  	 & \OII   &  6721.36   &  23.44   \\
  7000   & \CII   &  7231.34   &  16.33   \\
  	 & \CII   &  7236.42   &  16.33   \\
	 & \OII   &  7320.66   &   3.32   \\
	 & \OII   &  7331.30   &   3.33   \\
\hline
\end{tabular}
\end{table}

However, it is possible that the \OII\ levels are excited by non-thermal
processes, just like \HeI\ lines in SNe\,Ib (which actually come from lower
levels with smaller excitation potentials). We mimicked this by increasing the
population of the excited states of \OII\  (and \HeI) by values comparable to
values for SNe\,Ib \citep{lucy91, hachinger12} and SNe\,IIb \citep{mazzali09}.
We applied a single value for each epoch, but the departure coefficient
increases with time, from values of $\sim 100$ early on up to $\sim 10^4$ at the
later epochs. This reflects the progressive increase of the penetration of
non-thermalised particles through the outer ejecta. This is sufficient to
produce \OII\ lines that match the observations quite well at rest-frame
wavelenghts between 3000 and 6000\,\AA. 

\CII\ lines are quite strong. A \CII\ line ($\lambda 4267$) is important in
shaping the absorption of \OII\ near 4100\,\AA\ and making it broad. An
interesting prediction of this model is that \CII\ 6578 and 7256\,\AA\ should be
strong (see Fig.~\ref{fig:13ajgUVopt}). This region falls outside the
observed optical range for iPTF13ajg.  The NIR spectra reported by
\citet{vreeswijk14} are very noisy.

The time series of pre-maximum spectra is characterized by an essentially
constant photospheric velocity (12250\,\kms). The luminosity increase leading to
maximum light results in a temperature that decreases only slowly as the ejecta
expand, which ensures that the spectra show very little sign of change in this
period, which covers about one month. This is quite unusual for a SN. 

Because of the parametrized treatment of non-thermal processes (we do not know
the properties of the non-thermal radiation field, which is unlikely to be the
result of partial thermalization of gamma-rays and positrons), we cannot
establish the masses of the elements affected with great accuracy. Still, the
composition appears to be dominated by helium, carbon and oxygen, the elements
that are expected to dominate in the core of a massive star, while heavier
elements have sub-solar abundances. 

Our density profile has a mass of $\sim 3$\,\Msun\ above the photosphere.
Although this produces reasonable conditions in the thermal part of the
spectrum, this value should only be taken as indicative. The corresponding
kinetic energy of expansion is $\sim 3 \times 10^{52}$\,erg. This is quite
large, and is dictated by the need to have a steep gradient in density at
velocities of 10-15,000\,\kms\ in order to guarantee that the photosphere
recedes only slowly with time (in practice it corresponds to 3\,\Msun\ of
material moving at 12000\,\kms). A large mass is also required by the long
duration of the light curve. If a shorter risetime is used, a smaller mass and
energy are obtained.

\begin{figure}
 \includegraphics[width=89mm]{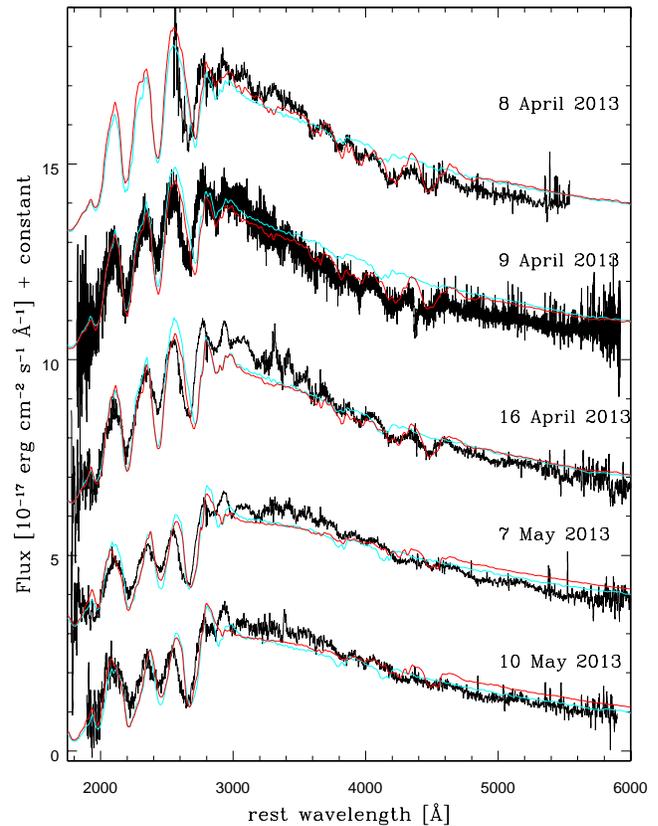}
\caption{The pre- and near-maximum spectra of iPTF13ajg, compared to synthetic 
spectra. One of the model series (cyan) does not include an increase in the 
occupation numbers of \OII\ and \HeI\ to account for non-thermal effects, while 
the other (red) does.}
\label{fig:13ajg_premax}
\end{figure}

Fig.~\ref{fig:13ajg_premax} shows the series of spectra before and
near maximum with their corresponding models, both without and with
enhanced excitation, to mimick non-thermal effects. The main
parameters of all the models are summarised in Table~\ref{tab:13ajg}.

\begin{figure}
 \includegraphics[width=89mm]{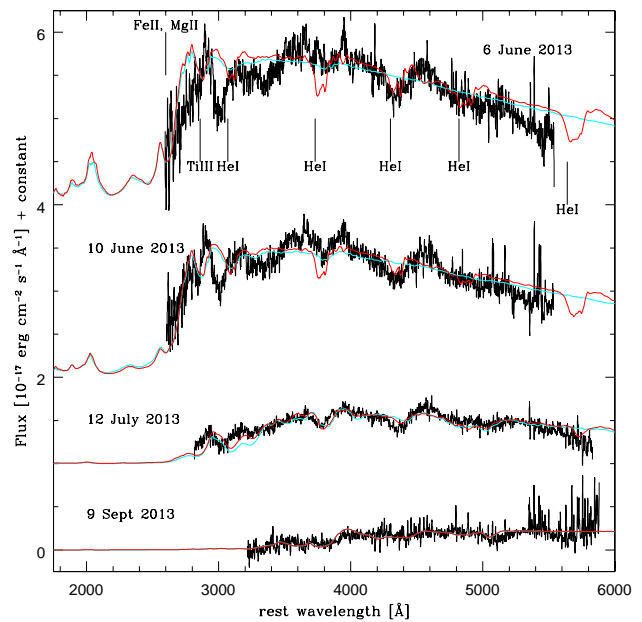}
 \caption{As Fig.\ref{fig:13ajg_premax}, but for the post-maximum
   spectra of iPTF13ajg.}
\label{fig:13ajg_postmax}
\end{figure}

\begin{table*}
\caption{Parameters of the models for PTF09cnd}
\label{tab:09cnd}
\centering
\begin{tabular}{ccccccccccccccccc}
Date & Epoch & Lum     & v$_{ph}$ & T$_B$ & He   & C    & O     & Ne   & Mg   & Si   & S    & Ca   & Ti   & Fe	& Co   & Ni  \\
 & [rf days] & erg s$^{-1}$ & km s$^{-1}$ & K &  \multicolumn{11}{c}{relative abundances}    \\
\hline
2009 Aug 12 &  18 & 43.22 & 15250 & 15600 & 0.42 & 0.10 & 0.27 & 0.02 & 0.08 & 0.10 & 5e-4 & 8e-7 & 5e-5 & 6e-5 & 2e-5 & 1e-5 \\
2009 Aug 25 &  29 & 43.48 & 14200 & 14400 & 0.44 & 0.11 & 0.30 & 0.02 & 0.06 & 0.06 & 6e-4 & 8e-7 & 6e-5 & 7e-5 & 2e-5 & 1e-5 \\
2010 Feb 11 & 164 & 43.04 &  3500 &  8200 & 0.46 & 0.11 & 0.40 & 0.02 & 5e-3 & 5e-4 & 5e-4 & 7e-7 & 1e-6 & 1e-4 & 1e-5 & 1e-5 \\
\hline
\end{tabular}
\end{table*}

\subsection{Post-Maximum epochs}
\label{sec:post-max}

After maximum light, the spectra of iPTF13ajg begin to evolve. The flux peak
progressively shifts to the red, which resembles SNe\,Ib/c. The \OII\ lines are
no longer very strong, and lines of lower ionization species become visible. No
rest-frame UV information is available, which limits our ability to constrain
the metal content in the regions where the spectra form \citep[see,
\eg][]{mazzali14b}. Observed and synthetic spectra are compared in
Fig.\,\ref{fig:13ajg_postmax}.

The decreasing luminosity is now accompanied by a decreasing photospheric
velocity, so that the mass above the photosphere becomes quite large at
progressively later epochs. We do not require any significant change in the
abundances, although the fits are not as good as at pre-maximum epochs (see
Fig.\,\ref{fig:13ajg_postmax}), and thus we are less confident about these
results.  \HeI\ lines now become visible, and a parametrised increase of the
level populations to take non-thermal heating into account is sufficient to
reproduce them. This was also used before maximum, but because most He was
ionized (\HeII) \HeI\ lines were never strong. Based on its post-maximum
spectra, iPTF13ajg should be classified as a SN\,Ib. It is normal for \HeI\
lines to appear late in SNe\,Ib, as the result of the delayed onset of
non-thermal effects on \HeI\ in a cool gas caused by the opacity encountered by
non-thermalized decay products as they travel in the SN ejecta \citep{lucy91}.
In the case of  iPTF13ajg, however, the late appearance of the \HeI\ lines is
also due to the late cooling and recombination of \HeII. On the other hand, the
ionization of oxygen is now too low for \OII\ lines ever to become visible.

The synthetic spectra do not match the observed spectra as well as they do at
epochs before maximum light. This may  partly be the consequence of our
ignorance of the flux level in the UV. A strong absorption near 3000\,\AA\ is
not reproduced. This feature is not often seen in SLSNe, but it seems to be
present in LSQS12dlf \citep{nicholl14}.

The trend for a cooler spectrum and decreasing velocities continues after
maximum. Lines of singly ionized species become more and more dominant as the
temperature decreases. Flux blocking in the (unobserved) UV is almost entirely
due to iron-group species. At the latest epochs, strong \CaII\,IR triplet,
\CI\ 9062, 9094, and \OI\ 7774 are predicted in the redder part of the visible
range, which is also not observed. This is in line with the spectra of other
SLSNe \citep[see][]{nicholl14}. 

As the photospheric velocity decreases with time, we are now exploring deeper
regions of the ejecta. Because of the steep density distribution that we
adopted, the enclosed mass increases rapidly with depth. In the final model
$\sim 20$\,\Msun\ of material contribute to the spectrum. This is necessary
because at an epoch of $\sim 140$ days the spectra still show significant
absorption.

\begin{figure}
 \includegraphics[width=89mm]{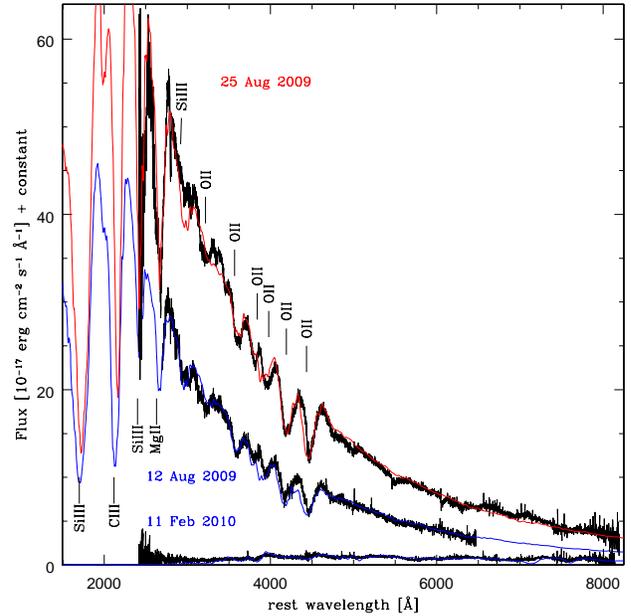}
\caption{The spectra of PTF09cnd (black), compared to synthetic spectra.}
\label{fig:09cnd}
\end{figure}

\section{Line-rich SLSNe: PTF09cnd }
\label{sec:09cnd}

The SLSN PTF09cnd \citep{quimby11}, discovered by the Palomar Transient Factory
(PTF), has coverage of both the pre-and post-maximum epochs. We use the original
classification spectrum taken at the 4.2-m William Herschel Telescope (WHT) with
the Intermediate dispersion Spectrograph and Imaging System (ISIS), as well as
two later spectra taken at the 10-m Keck-I telescope with the Low Resolution
Imaging Spectrograph (LRIS) \citep[see][for details on the spectra]{quimby11}.
Because of the relatively low redshift ($z = 0.258$), data only cover the UV
redwards of 2500\,\AA. PTF09cnd was significantly less luminous than iPTF13ajg,
and reached maximum more rapidly, but its spectra resemble those of iPTF13ajg,
showing the \OII\ line series in the optical prior to maximum and a few strong
metal lines in the only partially observed near-UV.  The combination of lower
luminosities and smaller epochs leads to a similar spectra evolution. In the
final spectrum \HeI\ lines are strong, so that PTF09cnd would also be classified
as a SN\,Ib.

\begin{table*}
\caption{Parameters of the models for SNLS-06D4eu}
\label{tab:06D4}
\centering
\begin{tabular}{cccccccccccccccccc}
Date & Epoch &     Lum & v$_{ph}$ & T$_B$ & He   & C   & O    & Ne   & Mg   & Al   &  Si  & S    & Ca   & Ti   & Fe   & Co   & Ni \\
 & [rf days] & erg s$^{-1}$ & km s$^{-1}$ & K &  &     &      &      &      &      &	  &	 &	&      &      &      & \\
\hline
2006 Aug 31 & 14 & 44.32 & 16500 & 18200 & 0.30 & 2e-3 & 0.43 & 0.02 & 0.10 & 2e-4 & 0.12 & 0.02 & 1e-5 & 5e-5 & 6e-3 & 1e-4 & 4e-5 \\
\hline
\end{tabular}
\end{table*}

We modelled PTF09cnd with the same background structure as iPTF13ajg, including
non-thermal excitation. A distance modulus $\mu = 40.55$\,mag was used.  The two
August spectra are both before maximum, which occurred towards the end of
August. Despite the lower luminosity, our fits indicate that the spectra are
shaped by the same lines which were identified in iPTF13ajg, although the UV is
only partially explored. The spectra can be reproduced using a smaller mass (7
-- 15\,\Msun), such that the pre-maximum spectra are shaped by only $\sim
0.5$\,\Msun\ of ejecta. As we mentioned above, the total mass cannot be
determined with great accuracy in the absence of nebular spectra. The mass above
the photosphere is subject to uncertainties in our estimated date of explosion
and non-thermal departure coefficients. The kinetic energy is 7 -- 15
$10^{51}$\,erg.  The main results are reported in Table~\ref{tab:09cnd}.
Synthetic spectra are compared to the observed ones in Fig.~\ref{fig:09cnd}. In
the figure, spectra are shown with their real flux level. Because of the very
rapid rise to maximum, the temperature in the earlier spectra is actually higher
than in iPTF13ajg. The spectra before maximum show very little evolution. The
line velocity also remains practically constant. Only much later, in the
post-maximum phase, do spectra turn red. At this point non-thermally excited
\OII\ lines are no longer visible.

Because of the limited UV coverage the determination of the metal content is
less reliable, but results seem to indicate a low metallicity, as in iPTF13ajg
or lower. One significant difference is the weakness of the \CII\ lines: the
lines between 3000 and 4000\,\AA\ are weak, while those between 6000 and
7000\,\AA\ are absent. Although this is partly due to the higher temperature,
which almost completely suppresses \CII, we still need to use a significantly
lower abundance than in iPTF13ajg. This is also useful to keep several UV lines,
and in particular the one near 2200\,\AA, which is dominated by \CIII\
2297\,\AA, from becoming too strong, creating too much fluorescent flux to the
red. We also need a much lower Fe abundance than in iPTF13ajg, again in order to
prevent very strong UV lines. None of these lines are covered by the spectra, so
these estimates are highly uncertain.

\section{Line-rich SLSNe: SNLS-06D4eu }
\label{sec:06D4eu}

\begin{figure}
 \includegraphics[width=89mm]{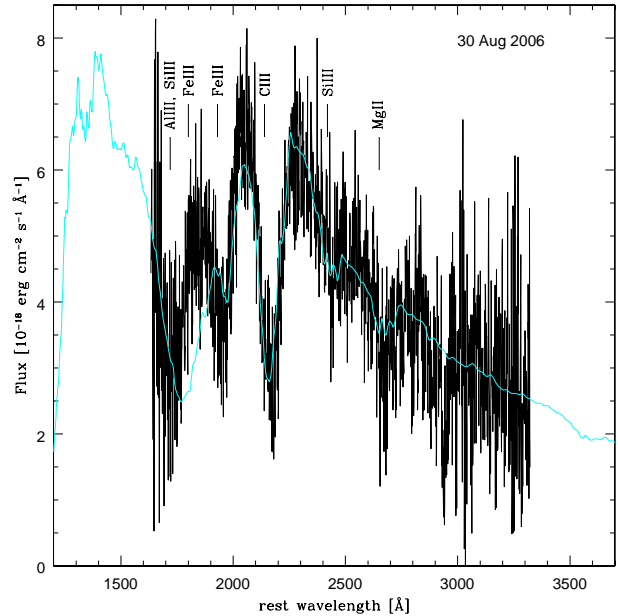}
 \caption{The near-maximum light spectrum of SNLS-06D4eu (black),
   compared to a synthetic spectrum (blue).}
\label{fig:06D4eu}
\end{figure}

The final line-rich SLSN that we consider is SNLS-06D4eu, the furthest
spectoscopically confirmed SLSN at the time of writing. The event was discovered
by the Supernova Legacy Survey \citep[SNLS;][]{perrett10} at a redshift
$z=1.588$ \citep{howell13}, confirmed using the European Southern Observatory
(ESO) Very Large Telescope (VLT) with the FORS1 instrument. It has a similar
luminosity as iPTF13ajg.  Because of the high redshift, the ground-based
spectrum only samples the rest-frame UV, reaching very short wavelengths. Since
the SN was optically faint and it was not spectrally identified at the time of
observation, only one spectrum is available, at an epoch close to maximum.  This
is shown in Fig.~\ref{fig:06D4eu} together with a synthetic spectrum computed
using a distance modulus $\mu = 45.34$\,mag.

The model that was used to compute the synthetic spectrum was again the same as
in the previous cases. The SN rose to peak relatively rapidly, so that the epoch
of spectrum is estimated to be only about 14 rest-frame days after explosion. 
This indicates a relatively low mass, and so we used the same mass as for
PTF09cnd ($\sim 15$\,\Msun). The rapid rise and the high luminosity require a
high velocity in order for the temperature and the ionization to reach the
observed values.  SNLS-06D4eu has the highest velocity of any of the line-rich
SLSNe analysed here. Only the GRB/SN\,2011kl reaches higher velocities (see next
section). We used $v \sim 16500$\,\kms, yielding a temperature near the
photosphere of $\sim 18000$\,K, the highest temperature of any of the spectra
modelled in this paper. The mass above the photosphere is only slightly less
than in the case of PTF09cnd. The high temperature causes the spectrum to have
an even bluer turnover (near 2000\,\AA) than previous cases ($\sim 2800$\,\AA).
Only $\sim 0.25$\,\Msun\ of material contribute to the spectrum.  Model
parameters are given in Table~\ref{tab:06D4}.

Beyond 2000\,\AA\ the spectrum is similar to those of other SLSNe, showing lines
of \CIII, \FeIII, \SiIII, \MgII.  The line near 2200\,\AA\ can be reproduced as
\CIII\ 2297\,\AA, but a very low carbon abundance is required to reproduce the
line strength, even lower than in the case of PTF09cnd.  \CII\ lines are very
weak in both PTF09cnd and SNLS-06D4eu, because the average ionization is much
higher than what would favour the formation of these lines. This is a
consequence of the high temperature. The lack of \CII\ lines may indeed be a
common feature of SLSNe at very early epochs. At the same time, a high iron
abundance is needed to reproduce the feature near 1800\,\AA\ as \FeIII. This is
in contrast with our model for PTF09cnd.  If this is correct, this suggests that
the ejecta of SNLS-06D4eu were nuclearly processed, so that it is not possible
to determine a metallicity. Silicon also has a high abundance, as indicated by
the strength of the line near 2400\,\AA, which is reproduced as \SiIII\
2542\,\AA, although the dominant ionization stage of silicon is \SiIV. 

\begin{table*}
\caption{Parameters of the models for GRB/SN}
\label{tab:GRB}
\centering
\begin{tabular}{ccccccccccccccccc}
Date & Epoch & Lum     & v$_{ph}$ & T$_B$ & He   & C   & O    & Ne   & Mg   &  Si  & S    & Ca   & Ti   & Fe   & Co   & Ni \\
 & [rf days] & erg s$^{-1}$ & km s$^{-1}$ & K &  &     &      &      &      &  	   &	  &	 &      &      &      & \\
\hline
2011 Dec 29 & 12 & 43.37 & 19500 & 10000 & 0.14 & 0.30 & 0.53 & 0.02 & 5e-5 & 0.01 & 5e-3 & 2e-4 & 1e-4 & 6e-4 & 1e-4 & 1e-5 \\
\hline
\end{tabular}
\end{table*}

Bluewards of 2000\,\AA, \FeIII\ lines are important, but the strong
absorption near 1800\,\AA\ does not seem compatible with \FeIII.  It
is better reproduced by the resonance line \AlIII\ 1854\,\AA. This
line is normally seen in the spectra of late-O stars \citep{walborn85}
as well as in B-stars \citep{rountree93}, including interacting
binaries \citep{mazzali87,mazzali92}. It has not -- to our knowledge
-- been individually resolved in type I SN spectra, as line blocking
by Fe lines is usually too strong \citep[][ identified this line in
the spectrum of SN\,1987A]{ml90}. The resonance line \SiIII\
1892\,\AA\ also contributes to this line. Many lines of doubly ionized
ions are visible; the singly ionized species are weak due to the
high temperature.  It should be noted that \FeIII\ lines do not have
the same predicted ratio in the synthetic spectrum as the observed
features. In particular, the synthetic absorption near 1700\,\AA\
appears too red because it is affected by \FeIII\ 1895, 1914,
1926\,\AA, which is necessary in order to form the line near
1950\,\AA, which is \FeIII\ 2068, 2079 \AA. The \MgII\ resonance line
(2795, 2805\,\AA) is visible near 2600\,\AA\ but is weak, while
\SiIII\ 2542 is responsible for the weak absorption near 2400\,\AA.

Since the spectra do not cover the rest-frame optical we cannot determine
whether \OII\ lines are as strong as in other SLSNe. However, the higher
ionization degree of this spectrum means that strong NLTE effects are not
required to cause the \OII\ lines to appear.

\section{Line-poor SLSNe: GRB/SN 2011kl }
\label{sec:SN2011kl}

While most known SLSNe show similar spectra, recently a SN has been discovered
in association with an ultra-long gamma-ray burst (ULGRB), GRB111209A at a
redshift $z = 0.67$. The SN shows many characteristics of SLSNe. SN\,2011kl was
much more luminous than classical GRB/SNe, but was not quite as luminous as
SLSNe \citep{greiner15}. Its light curve evolved over just a few weeks, as do
GRB/SNe, but its spectrum (available only near maximum) did not show any of the
broad lines that characterize GRB/SNe \citep[see, \eg][]{iwamoto98} or
broad-lined SNe\,Ic \citep[see, \eg][]{m13}. The spectrum was blue, turning over
only at $\sim 3000$\,\AA, like SLSNe.  Unlike SLSNe, it was rather featureless,
with only one strong and rather broad absorption line, near 2800\,\AA. 

In \citet{greiner15} we adopted a model similar to the one used here to
reproduce the spectrum of SN2011\,kl. Here we highlight some of the features of
that model, and contrast them to those of SLSNe.  A distance modulus $\mu =
43.01$\,mag was used.  Model parameters are given in Table~\ref{tab:GRB}.

\begin{figure}
 \includegraphics[width=89mm]{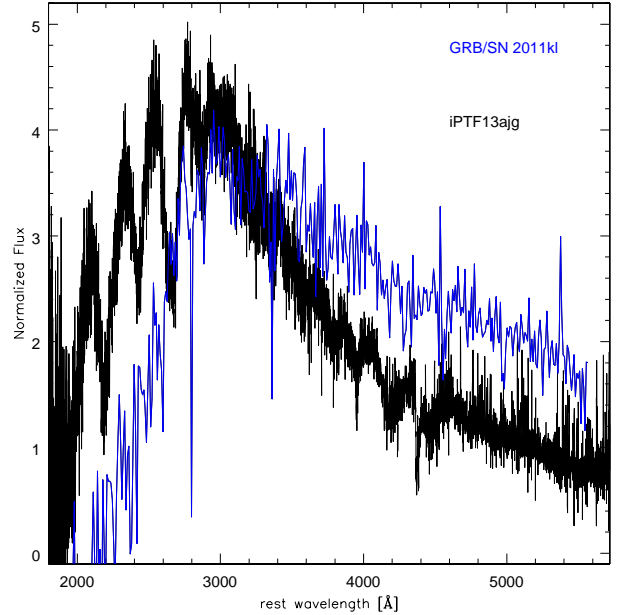}
\caption{The maximum-light spectrum of GRB/SN 2011kl (blue line) compared to 
that of iPF13ajg (black line).}
\label{fig:GRBajg}
\end{figure}

In Fig.~\ref{fig:GRBajg} we compare the spectrum of SN\,2011kl with that of
iPTF13ajg. The similarity is clear. If the line seen near 2900\AA\ is that same
as the line which is identified as \TiIII\ in iPTF13ajg, then the velocity must
be higher in SN\,2011kl, as the blueshift is larger. We adopted for SN\,2011kl a
similar density distribution as for iPTF13ajg, but set a photospheric velocity
of $\sim 20000$\,\kms, as opposed to $\sim 12000$\,\kms.

The spectrum of SN\,2011kl appears somewhat cooler, which is the result of the
lower luminosity and the higher expansion velocity. The high photospheric
velocity means that line blending is stronger in SN\,2011kl. Line blending is
particularly strong in the UV, where many lines are present which merge causing
significant line blanketing.  This is seen also in GRB/SNe, which are
characterised by a high velocity. The metal content is quite similar in the two
events. The spectrum does not show strong \OII\ lines, suggesting that
non-thermal effects are not strong in SN\,2011kl.

\begin{figure}
 \includegraphics[width=89mm]{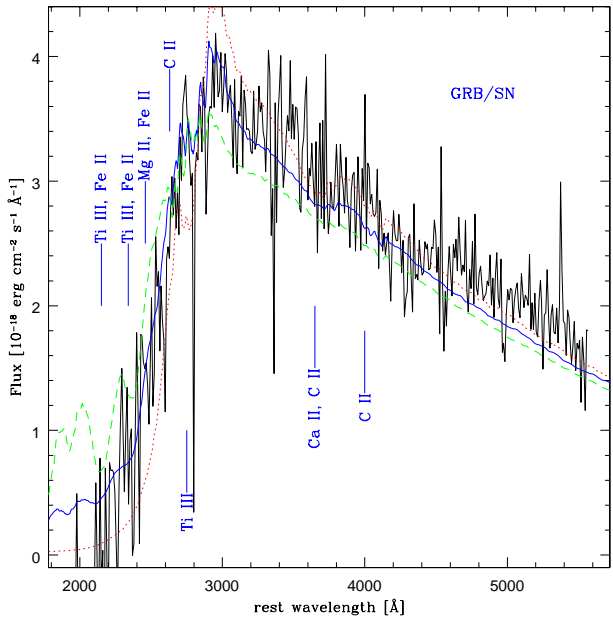}
\caption{The maximum-light spectrum of GRB/SN 2011kl compared to synthetic 
spectra. The blue (solid) line shows a model with metallicity 
$Z \sim 1/4 Z_{\odot}$. The red (dotted) line shows a model with metallicity
reduced by a factor of 3 ($Z \sim 1/12 Z_{\odot}$). The green (dashed) line 
shows a model with metallicity increased by a factor of 3 
($Z \sim 3/4 Z_{\odot}$). 
The strong narrow absorption line near 2800\,\AA\ is the \MgII\ resonance 
line from the host galaxy.}
\label{fig:GRBSNmodel}
\end{figure}

In order to set the level of the UV flux, the mass above the photosphere and the
abundance of metals can be adjusted. We find that a mass of $\sim 0.3$\,\Msun\
above 19000\,\kms\ leads to the best convergence.  Since only one spectrum is
available, it is not possible to determine whether the density profile is the
same at lower velocities.  The rapid evolution of the light curve however
suggests a smaller mass \citep[$\sim2.5$\,\Msun,][]{greiner15}, and a higher
ratio of kinetic energy to ejected mass.  This is similar to the case of GRB/SNe
when compared to normal type Ib/c SNe.  Keeping the composition similar to the
model of iPTF13ajg, and modifying only the ratio of metals versus carbon and
oxygen, we find that a metal content of $\sim 1/4$ solar is required for
SN2011\,kl, which is in line with the metallicity estimated for the host galaxy
\citep{greiner15}. Additionally, the temperature is lower because of the lower
luminosity. Line blanketing and the lower temperature depress the UV spectrum
shortwards of 3000\,\AA\ compared to iPTF13ajg \ref{fig:GRBajg}. Our synthetic
spectrum is compared to the observed one in Fig.\,\ref{fig:GRBSNmodel}. 

We tested different metal abundances by modifying the abaundances of all
elements heavier than Neon. The results are shown in \ref{fig:GRBajg}.  At lower
metal content line blending is reduced and some of the individual line features
begin to appear. Also, the UV flux shortwards of 3000\,\AA\ is too high, so less
radiation is reprocessed to the red and the flux above 3000\,\AA\ is too low.
Conversely, if the metal content is too high, the flux shortwards of 3000\,\AA\
is suppressed, and the excessive reprocessing least to a strong peak near
3000\,\AA.

Interestingly the metallicity used for SN\,2011kl is similar to that in
iPTF13ajg. Several papers on SLSNe estimated the star-formation rate in the host
galaxy, while it is less common to find an estimate of the metallicity. In the
case of SN\,2006oz, \citet{leloudas12} quote a value $Z/Z_{\odot} \sim 0.17
^{+0.18}_{-0.08}$, which is in line with our estimates in SLSNe. So our models
may be used to estimate the metallicity of the progenitor, as in the case of
other SNe \citep{mazzali14b}.

\section{Discussion}
\label{sec:disc}

Our models show that the spectra of SLSNe can be reproduced almost
self-consistently by making just a few assumptions, which are necessary to
reproduce some of the peculiar features of these spectra. These observed
features and their implications are listed below.

\begin{itemize}

\item The spectral lines do not evolve in velocity until well after maximum:
  this requires a steep density gradient, such that the decrease in density with
  radius is faster than that with time. We adopted a density structure
  characterised by $\rho \propto r^{-7}$. This is typical of low-energy 1D
  explosions of compact progenitors \citep[\eg][]{sauer06}, but is not normally
  seen in high-energy SNe Ib/c, and it may suggest that SLSNe are overall more
  spherically symmetric than SNe\,Ib/c with a high \KE/\Mej\ \citep{leloudas15}. 
  SLSNe seem to have a low \KE/\Mej$\sim 1 [10^{51}$\,erg \Msun$^{-1}$], as 
  shown in Table \ref{tab:props}.

\item The line velocities are not particularly high. This can be
  simulated by adjusting the density. The steep density profile also
  means that the lines are not very broad.

\item The light curves of many SLSNe are broad. This requires a large
  ejected mass, independent of the powering mechanism.

\item Spectral lines from very highly-excited states are observed. Even at the
  high luminosities of SLSNe, lines such as \OII, which are strong in the
  optical before maximum, require non-thermal excitation processes. \HeI\ lines
  are affected as well, but only appear at later epochs, when the temperature is
  lower and He is not mostly \HeII.

\item The metal lines are weak, and the spectra are blue. This requires a fairly
  low metal abundance. In the case of iPTF13ajg we find a metallicity of $\sim
  1/4$ solar. This only produces weak iron lines, and allows the spectrum to
  turn over at $\sim 3000$\,\AA, compared to $\sim 4000$\,\AA\ in other
  SNe\,Ib/c. SN\,2011kl has a similar metallicity, but line blanketing depresses
  the UV flux shortwards of $\sim 3000$\,\AA.

\item The composition of the ejecta that contribute to spectrum formation is
  consistent with the CO cores of massive stars, indicating a common origin of
  SNe Ib/c and SLSNe. The latter appear to have larger masses on average, but it
  may well be that such events are rare and only the brightest end has been
  uncovered so far.  The case of the GRB/SN\,2011kl, which was not as luminous
  as the rest of the SLSNe but was discovered only thanks to the GRB, indicates
  that this may indeed be the case. SNLS-06D4eu may show signs of nuclearly
  processed material.

\item The fact that the spectrum can be reproduced using the full SN luminosity
including the depth of the lines suggests that these SNe are not energised by
interaction with an external H- and He-poor CSM.

\end{itemize}

While these are general results, it is difficult to be precise about the mass
and energy of the explosion. Our models (see Table \ref{tab:props}) require
large masses, but do not explore the inner layers because there are no very
late-time data. Therefore the mass values given in Table \ref{tab:props} should
be regarded as estimates only. The kinetic energy of our models range from 0.4
to $\sim 5 \times 10^{52}$\,erg. This range is typical of GRB/SNe. The ratio
\KE/\Mej\ is not far from 1 [$10^{51} {\rm erg}\, \Msun^{-1}$], ensuring that
lines are not too broad, except for GRB/SN\,2011kl, which reaches \KE/\Mej\
$\approx 2$ and has blended lines.  Determining the mass of the inner region has
to rely on parametrised fits of the light curve, as we do not have nebular
spectra. This has been done in many papers \citep[e.g.,][]{nicholl15}.
Considering the uncertainties (risetime, structure of the inner ejecta), results
are in general agreement with ours. 

\begin{table}
\caption{Inferred ejecta masses and energies of SLSNe}
\label{tab:props}
\centering
\begin{tabular}{ccc}
SN &  Mass & \KE    \\
   & \Msun & erg    \\
\hline
iPTF13ajg      & $25-50$  & $2.5-5.0 \times 10^{52}$ \\
 PTF09cnd      &  $7-15$  & $7 - 15 \times 10^{51}$ \\
 SNLS-06D4eu   &  $7-15$  & $7 - 15 \times 10^{51}$ \\
SN\,2011kl     &  $2-3$   & $5 - 8  \times 10^{51}$ \\
\hline
\end{tabular}
\end{table}

Finally, are normal SNe\,Ib/c, GRB/SNe, ULGRB/SNe and SLSNe related? They
probably all originate from stripped cores of massive stars. In the case of
SNe\,Ib/c the light curve is dominated by \Nifs, and the energy seems to be
proportional to the mass and the \Nifs\ mass \citep{m13}. The inferred
progenitor mass rarely exceeds 25\,\Msun. They may be classical neutrino-driven
explosions, although the increased \Nifs\ mass and \KE\ may suggest an
additional contribution to the energetics of the event, such as the increasing
impact of a magnetar \citep{mazzali06}. GRB/SNe are more massive (35-50\,\Msun),
produce more \Nifs, but are also much more energetic. While the collapsar
mechanism is a serious possibility \citep{McFW99}, the fact that all
well-studied GRB/SNe are very similar and that the \KE\ is very nearly constant
and close to $2 \times 10^{52}$\,erg, the limiting energy of a magnetized
neutron star, suggests that magnetar energy may be the driving mechanism
\citep{mazzali14a}. In this case magnetar energy would need to be applied very
early on, so that it could contribute to nucleosynthesis as well as to
energising the explosion. In the case of the ULGRB/SN the GRB was so long that
it may not be compatible with typically assumed black hole accretion rates in
the Collapsar mechanism \citep{woosley93,McFW99,McFWooHeg03} unless a very large
mass is accreted. Thus the magnetar energy injection scenario may be an
appealing alternative. Finally, for classical SLSNe, the constancy of the line
velocity as well as the very large luminosity suggests again that magnetar
energy plays a major role in shaping the SN ejecta. In this case magnetar energy
may not be injected immediately and may not contribute to nucleosynthesis,
although it may add \KE.  The \KE\ deduced from our models is again in line with
the maximum magnetar energy.  The delayed  addition of magnetar energy may shock
the SN ejecta and produce X-rays which then give rise to the non-thermal
spectrum and the optical radiation which sustains the light curve for a much
longer time than GRB/SNe.  This process may also lead to the steep density
profile which is needed for the stationary lines, although this may just be
explained if the explosion was spherically symmetric, in which case magnetar
energy would act like a ``piston''. Part of the reason for the extended light
curve of SLSNe is also the large mass of the SN ejecta. 

The spectroscopic properties of non-interacting SLSNe reflect those of
SNe\,Ib/c. Although near peak most non-interacting SLSNe would be classified as
type Ic, He lines ofter appear after maximum, when the temperature drops, so
they should be re-classified as SNe\,Ib. The reason for the late appearance of
the He lines is however different in the two cases: in normal SNe\,Ib
non-thermal effects populating the excited levels of \HeI\ above their themal
values can only happen late, when non-thermal particles can penetrate through
the ejecta and deposit their energy in the He layer. In SLSNe, on the other
hand, the temperature near maximum, when  non-thermal effects are important for
\OII, is too high for \HeI\ lines, which become strong when He starts to
recombine. When they finally appear, \HeI\ lines are also subject to non-thermal
excitation effects.

This leads us to making another comment: if hydrogen was present in SLSNe it
would most likely not be visible near maximum because it would be almost
completely ionized. The lack of observed H lines near maximum in SLSNe suggests
that any H content should not be very large. SLSNe with a lot of H may belong to
the interacting class (similar to SNe\,IIn), while SLSNe of type IIP may not
exist because the progenitors of SLSNe may be too massive to retain the H 
envelope. However, it would not be surprising to see SLSNe showing H lines as
they cool down after maximum. These would then formally be ``SLSN-IIb'', making
the ``Type I'' classification inaccurate. H has recently been observed to be
present very close to the SN \citep{Yan2015}. It seems therefore dangerous to
classify SLSNe as stripped-envelope subtypes, or even as ``Type I'' or H-poor,
as long as we do not have a more complete temporal coverage of their properties.
We have adopted the definition SLSN-I in this paper for clarity, but we think
this is not a fully appropriate definition. We do however believe that
non-interacting SLSNe share the same properties as stripped-envelope SN, as
suggested by \citet{pastorello10}.

On the other hand, GRB production seems to be agnostic about whether the stellar
explosion turns out to be a SN\,Ib/c or a SLSN. If GRBs are beamed, we should
expect to find 2011kl-like SLSN without a GRB. SN\,2011kl was less luminous than
`normal' SLSNe, so such events may not have been discovered yet because they are
rare and too faint to be discovered optically at their typically large
distances. We also should expect that there is no discontinuity in luminosity
between SN\,2011kl and `typical' SLSNe. 

SLSNe are so luminous that they are unlikely to be driven by \Nifs, but at the
same time it is diffcult to determine just how much \Nifs\ they contain. If a
SLSN synthesised as much \Nifs\ as a GRB/SN this would go almost unnoticed
because the light curve is a lot brighter. It may be expected that, if and when
the contribution of whatever alternative power source fades, the SN may fall
back on the \Nifs\ light curve, as was the case for SN\,2005bf \citep{maeda07}. 
Apart from this particular case, this has never been observed, possibly because
it is difficult to follow SLSNe for a very long time. Magnetar solutions predict
a different late-time behaviour than \Nifs\ solutions \citep[see,
\eg][]{greiner15}, so it is possible that the real \Nifs\ contribution remains
hidden for a long time. Interaction with an outside CSM, on the other hand,
should terminate at some point, and the light curve fade rapidly. The fact that
this has not been observed is another argument against the outer CSM scenario.
Very late-time observations may shed light on this, as they did in the case of
the pair-instability candidate SN\,2007bi \citep{GalYam09}, where the emission
in Fe lines matches the expectations for the \Nifs\ content required to power
the peak of the light curve. An additional reason to think that SN\,2007bi and
similar events are not magnetar-driven SLSNe is the lack of the \OII\ line
series despite the O-rich composition, which suggest lack of a strong X-ray flux
despite the large oxygen mass inferred from the nebular spectra. Since
SN\,2007bi was O-rich, this means that X-rays from \Nifs\ should be less
powerful than X-rays from the shock.

We therefore may have a situation where prompt magnetar energy injection may
produce nucleosynthesis in GRB/SN and help drive the explosion. If the energy is
injected later, as would be the case for a lower magnetic field, it would not 
contribute to nucleosynthesis, but it may be converted into radiative energy.
This may be the case for SLSNe, but also for SN\,2005bf
\citep{tominaga05,maeda07}, which was the result of the explosion of a less
massive star. The late excitation of the \HeI\ lines in this case may well be a
consequence of the injection of X-rays from interaction with a magnetar wind,
since the mass of \Nifs\ synthesised was not very large \citep{maeda07}. It
cannot be ruled out that SLSNe commonly host ULGRBs. In this case the presence
of a He shell would probably not constitute a problem for the propagation of the
jet, because the engine is active for a very long time. 

On the other hand, non-interacting SLSNe may be divided in
magnetar-driven and \Nifs-driven, and the latter may be identified as
PISN or massive and energetic core-collapses \citep{moriya10}. The
presence or lack of the \OII\ line series may be a way to discriminate
between them, as are the behaviour of the light curve and the
late-time spectrum.

\section*{ACKNOWLEDGEMENTS}
We are thankful to Paul Vreeswijk and Avishay Gal-Yam for making the
spectra of iPTF13ajg available to us, and to Ken'ichi Nomoto for
useful discussions. MS acknowledges support from the Royal Society and
EU/FP7-ERC grant no [615929]. The William Herschel Telescope is
operated on the island of La Palma by the Isaac Newton Group in the
Spanish Observatorio del Roque de los Muchachos of the Instituto de
Astrofísica de Canarias.

\end{document}